\documentclass[twocolumn,showpacs,superscriptaddress,prl,citeautoscript,amsmath,amssymb,letterpaper]{revtex4-2}
\usepackage{graphicx}
\usepackage{multirow}
\usepackage{color}
\usepackage{bm}
\usepackage{times}
\usepackage{amsmath,bm,amsfonts}
\usepackage{dcolumn}
\usepackage{graphicx}
\usepackage{latexsym}
\usepackage{hhline}
\usepackage{comment}

\usepackage{braket}
\usepackage{bbold}
\usepackage{multirow}

\def\ket#1{|#1\rangle }

\def\d{\partial}

\renewcommand{\thetable}{\arabic{table}}

\begin{document}
\title{
Topological Circular Dichroism in Chiral Multifold Semimetals
}
\author{Junyeong \surname{Ahn}}
\email{junyeongahn@fas.harvard.edu}
\affiliation{Department of Physics, Harvard University, Cambridge, MA 02138, USA}	

\author{Barun \surname{Ghosh}}
\email{b.ghosh@northeastern.edu}
\affiliation{Department of Physics, Northeastern University, Boston, MA 02115, USA}

\begin{abstract}
Uncovering the physical contents of the nontrivial topology of quantum states is a critical problem in condensed matter physics.
Here, we study the topological circular dichroism in chiral semimetals using linear response theory and first-principles calculations.
We show that, when the low-energy spectrum respects emergent SO(3) rotational symmetry, topological circular dichroism is forbidden for Weyl fermions, and thus is unique to chiral multifold fermions.
This is a result of the selection rule that is imposed by the emergent symmetry under the combination of particle-hole conjugation and spatial inversion.
Using first-principles calculations, we predict that topological circular dichroism occurs in CoSi for photon energy below about 0.2 eV.
Our work demonstrates the existence of a response property of unconventional fermions that is fundamentally different from the response of Dirac and Weyl fermions, motivating further study to uncover other unique responses.
\end{abstract}

\date{\today}

\maketitle

{\it Introduction.---}
The interaction between chiral materials and circularly polarized light is a topic of broad interest in fundamental sciences~\cite{huck1996dynamic,bailey1998circular,feringa1999absolute,malashevich2010band,zhong2016gyrotropic,xu2020spontaneous,de2017quantized,orenstein2021topology,ma2021topology,PhysRevLett.129.227401}.
Because chiral materials have a definite left- or right-handed crystalline structure, they respond differently to the left and right circularly polarized light.
Natural optical activity (i.e.,  optical rotation and circular dichroism with time-reversal symmetry) and the circular photogalvanic effect are such phenomena due to the light-helicity dependence in the refractive index and DC photocurrent, respectively.

The quantization of the circular photogalvanic effect in chiral topological semimetals has gained attention recently~\cite{de2017quantized,flicker2018chiral,rees2020helicity,ni2021giant,orenstein2021topology,ma2021topology}.
In three-dimensional chiral crystals, a band-crossing point carries a quantized magnetic monopole charge in momentum space, which is the Chern number~\cite{bradlyn2016beyond,chang2018topological}.
While the magnetic monopoles appear in pairs in the Brillouin zone by the fermion doubling theorem~\cite{nielsen1981no}, monopole and anti-monopole are not at the same energy, in general, because there is no symmetry to relate them in chiral crystals.
The uncompensated monopole charge of a chiral fermion near the Fermi level can manifest through physical responses.
The quantized circular photogalvanic effect is a rare example of topological optical responses originating from the monopole charge of a chiral fermion.

\begin{figure}[b]
\includegraphics[width=0.48\textwidth]{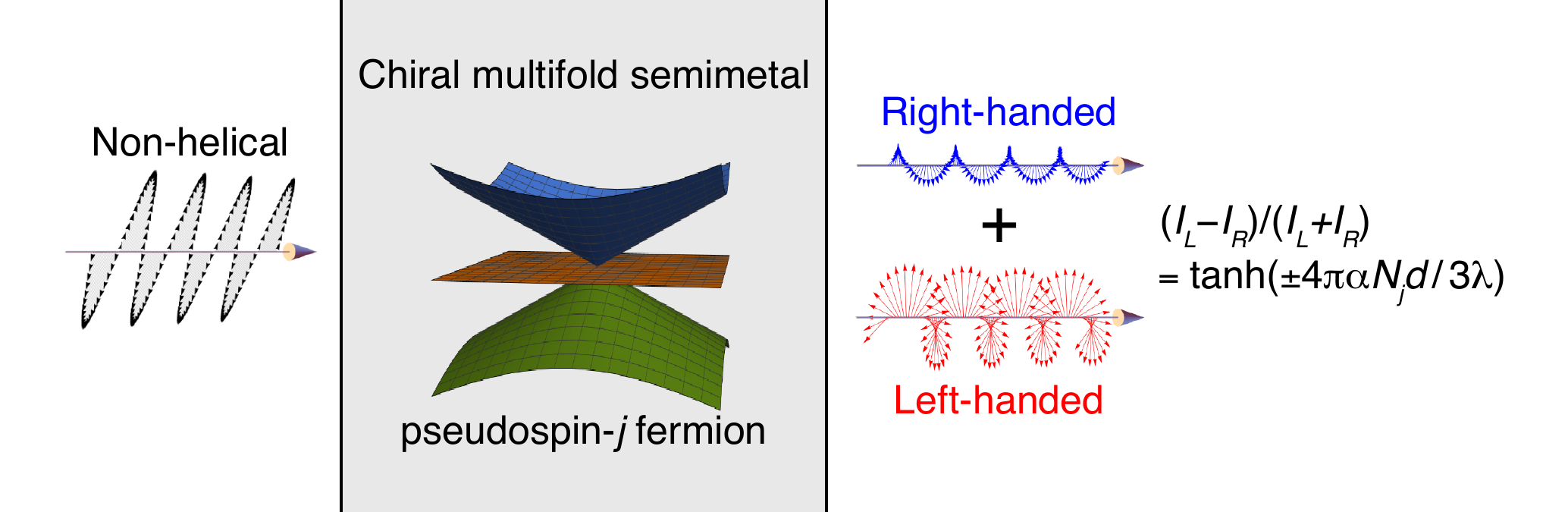}
\caption{
Topological circular dichroism by a chiral multifold semimetal hosting a pseudospin-$j$ fermion near the Fermi level.
$I_{L/R}$s are the transmitted intensity for the left ($L$) and right ($R$) handed light. $N_{1/2}=0$, $N_1=1$, and $N_{3/2}=3$.}
\label{fig:scheme}
\end{figure}

More recently, another topological optical phenomenon was discovered in chiral topological semimetals~\cite{sekh2022circular,mandal2023signatures}.
It was proposed that linearly dispersing chiral fermions show topological circular dichroism, where the helicity-dependent absorption of light is determined only by universal quantities, including fundamental constants and the ratio between the sample thickness and the light wavelength [Fig.~\ref{fig:scheme}].
While this discovery provides another exciting example of topological optical responses, the results in Refs.~\cite{sekh2022circular,mandal2023signatures} need further investigation because they were derived from physical arguments using Fermi's Golden rule without rigorous derivations.

In this Letter, we investigate topological circular dichroism in chiral topological semimetals using linear response theory and first-principles calculations.
Remarkably, we find that topological circular dichroism does not appear for Weyl fermions, which are chiral fermions with twofold degenerate band-crossing points, and is thus unique to chiral multifold fermions having three- or four-fold degenerate band-crossing points.
We also find differences in the magnitude and spectral range of the quantized response for chiral multifold fermions compared to the original proposal.
We show that these new features are mainly because of the selection rule imposed by the symmetry under the combination of particle-hole conjugation and spatial inversion.

Unlike the quantized circular photogalvanic effect, topological circular dichroism does not depend on the current relaxation time, which depends on materials.
Instead, the topological circular dichroism relies on isotropic linear dispersion.
To test our model analysis, we perform first-principles calculations of the circular dichroism for CoSi, a chiral threefold semimetal with good linear dispersions per spin degrees of freedom~\cite{chang2017unconventional,tang2017multiple,CoSi_arpes_2,sanchez2019topological,xu2020optical}.
The result agrees well with model analysis, showing approximate quantizations for photon energies below about 0.2 eV.

\begin{figure}[t]
\includegraphics[width=0.48\textwidth]{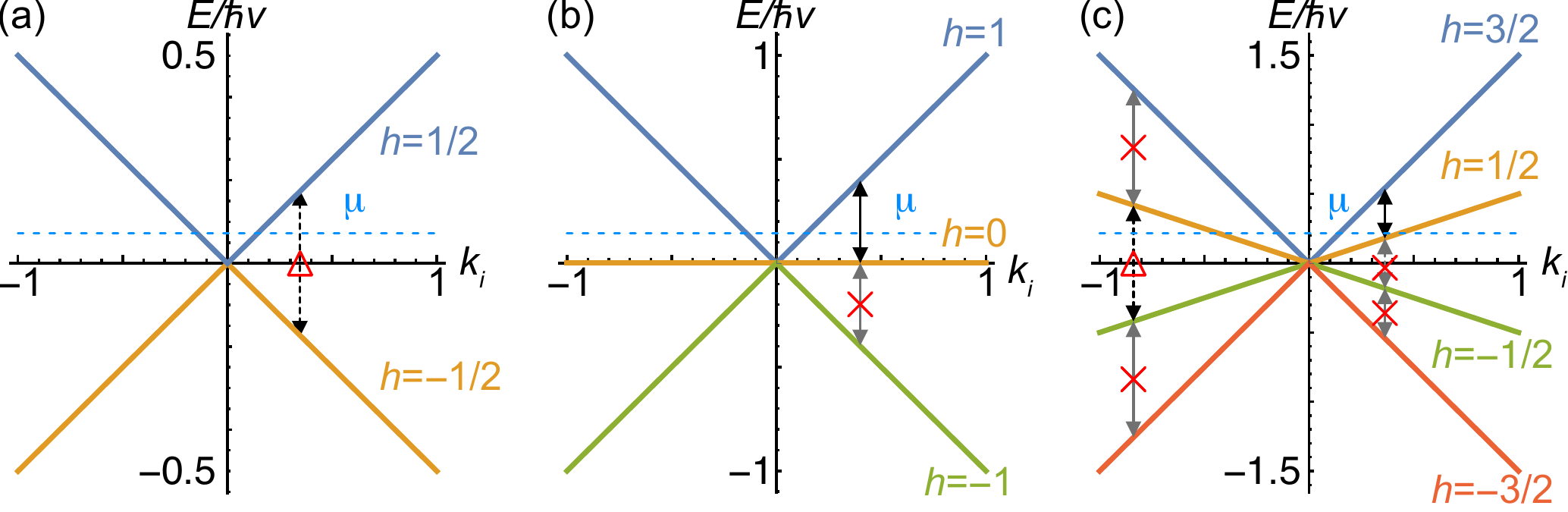}
\caption{
Band structure of pseudospin-$j$ fermions described by Eq.~\eqref{eq:model}.
(a) $j=1/2$.
(b) $j=1$.
(c) $j=3/2$.
The spectrum has the same shape along $k_{i=x,y,z}$ because of isotropy.
Arrows represent possible optical transition channels allowed by the selection rule due to isotropy~\cite{chang2017unconventional,sanchez2019linear}.
Optical transitions with red x marks are forbidden by the Pauli blocking with the chemical potential represented by the blue dashed line.
$CP$ symmetry further constrains that transitions between $h=\pm 1/2$ bands (arrows with red triangles) does not contribute to natural optical activity.
}
\label{fig:BS}
\end{figure}
{\it Isotropic k dot p model.---}
We first consider the model of isotropic chiral pseudospin-$j$ fermions in three dimensions~\cite{bradlyn2016beyond,tang2017multiple}.
\begin{align}
\label{eq:model}
H_0({\bf k})=-\mu+\chi v{\bf k}\cdot{\bf J},
\end{align}
where ${\bf k}$ is the wave vector, ${\bf J}$ is the pseudospin-$j$ operator satisfying the su(2) algebra because of isotropy.
The sign $\chi=\pm 1$ determines the chirality.
The energy eigenvalues are
\begin{align}
E_n({\bf k})=-\mu+\hbar v kh_n,
\end{align}
where the integer $h_n=-j,\hdots,j$ is the helicity quantum number [Fig.~\ref{fig:BS}].
The crossing point at ${\bf k}=0$ has $(2j+1)$-fold degeneracy.
The band with helicity $h$ carries the Chern number $c_{h}=-2\chi h$ on a closed surface that encloses the node (i.e., the magnetic monopole charge in momentum space defined by the Berry curvature), which serves as a topological charge of the spin-$j$ fermion.
We have a Weyl fermion for $j=1/2$ and a chiral multifold fermion for a higher $j$.
In this model, optical transitions occur between adjacent energy levels only because of an optical selection rule imposed by isotropy~\cite{chang2017unconventional,sanchez2019linear}:
for $m\ne n$, transition dipole moment $\braket{\psi_{m\bf k}|e\hat{\bf r}|\psi_{n\bf k}}\propto\braket{u_{m\bf k}|{\bf J}|u_{n\bf k}}
=0$ if $h_m\ne h_n\pm 1$.

Our model has symmetry $TH({\bf k})T^{-1}=H({\bf -
k})$ under effective time reversal $T$ that flips the pseudospin.
Therefore, the anomalous Hall effect is forbidden.
However, natural optical activity can arise from broken inversion symmetry.

Below we focus on isotropic spin-1 fermions because they are more relevant to real materials but consider spin-3/2 fermions as well for completeness.
In crystals, the topological protection of multifold fermions requires particular space group symmetries.
When spin-orbit coupling is negligible, space groups 195-199 and 207-214 combined with time reversal symmetry can protect isotropic threefold fermions~\cite{bradlyn2016beyond}.
A topologically stable isotropic threefold fermion can also appear in spin-orbit coupled antiferromagnets with type IV magnetic space groups $P_I$213(198.11) $P_I$4$_3$32(212.62) and $P_I$4$_1$32(213.66) ~\cite{cano2019multifold}.
On the other hand, a threefold fermion stabilized by other space group symmetries do not respect full isotropy in the low-energy limit~\cite{bradlyn2016beyond,cano2019multifold}.
Topologically protected spin-3/2 fermions does not have isotropy unless fine tuned~\cite{bradlyn2016beyond}.
Nevertheless, we do not exclude the possibility of a fine-tuned isotropic spin-3/2 fermion and consider both isotropic spin-1 and spin-3/2 fermions.

{\it Topological circular dichroism from natural optical activity.---}
In crystalline solids, natural optical activity is described by the part of the optical conductivity that is linear in photon momentum ${\bf q}$~\cite{malashevich2010band}.
Let us consider the expansion $\sigma_{ab}(\omega,{\bf q})=\sigma_{ab}(\omega)+\sigma_{abc}(\omega)q_c+O(q^2)$.
In our model, the refractive indices for light with left ($L$) and right ($R$) helicity are
\begin{align}
\label{eq:npm}
n_{L/R}=\sqrt{1+\chi_{xx}+(\mu_0c\sigma_{xyz}/2)^2}\pm \mu_0c\sigma_{xyz}/2,
\end{align}
where $\chi_{ab}=\sigma_{ab}(-i\epsilon_0\omega)^{-1}$ is the electric susceptibility, and the light helicity is defined by the sign of ${\bf q}\cdot i{\bf E}^*\times {\bf E}$.
For ${\bf q}=|{\bf q}|(0,0,1)$, $L$ and $R$ polarization vectors are respectively $\hat{L}=(1,-i,0)/\sqrt{2}$ and $\hat{R}=(1,i,0)/\sqrt{2}$.
Because of the isotropy in our model, $\chi_{xx}$ and $\sigma_{xyz}$ are the only nonvanishing tensor components.
The real and imaginary parts of circular birefringence $n_L-n_R=\mu_0c\sigma_{xyz}$ are responsible for the optical rotation and circular dichroism, respectively.

Natural optical activity has two contributions from the Fermi sea and the Fermi surface, respectively~\cite{malashevich2010band,zhong2016gyrotropic,pozo2023multipole}.
The formula for the Fermi sea part is~\cite{malashevich2010band}
\begin{align}
\label{eq:FermiSea}
\sigma_{abc}^0
&=\frac{e^2\omega}{\hbar}
\sum_{n,m}\int_{\bf k}
f_{nm}
\bigg[
\frac{{\rm Im}(r^a_{nm}B^{bc}_{mn}-r^b_{nm}B^{ac}_{mn})}{\omega_{mn}^2-\omega^2}\notag\\
&\qquad-\frac{(3\omega_{mn}^2-\omega^2)
{\rm Im}(r^{a}_{nm}r^b_{mn})
(v^c_{mm}+v^c_{nn})}{2(\omega_{mn}^2-\omega^2)^2}
\bigg],
\end{align}
where $\int_{\bf k}=\int_{\rm BZ} d^3k/(2\pi)^3$, $f_{nm}=f_n-f_m$ and $\hbar\omega_{mn}=\hbar\omega_m-\hbar\omega_n$ are the differences of the Fermi-Dirac distributions and energy eigenvalues, respectively, $v^i_{mn}=\braket{\psi_{m\bf k}|\hat{v}^i|\psi_{n\bf k}}$ and $r^j_{nm}=-iv^j_{nm}/\omega_{nm}$ are velocity and position matrix elements, $B^{ab}_{mn}=B^{{\rm orb},ab}_{mn}+B^{{\rm spin},ab}_{mn}$,
$B^{{\rm orb},ab}_{mn}=2^{-1}(\sum_{p;E_p\ne E_m}r^b_{mp}v^a_{pn}+\sum_{p;E_p\ne E_n}v^a_{mp}r^b_{pn})$, $B^{{\rm spin},ab}_{mn}=e^{-1}\epsilon_{abc}\braket{\psi_{m\bf k}|\hat{M}^{{\rm spin}, c}|\psi_{n\bf k}}$, and $\hat{\bf M}^{{\rm spin}}$ is the spin magnetic moment operator.
The spin magnetic moment does not contribute to the response in systems with negligible spin-orbit coupling; we discuss its effect in spin-orbit coupled systems below.
The Fermi surface part is given by~\cite{goswami2013chiral,zhong2016gyrotropic}
\begin{align}
\label{eq:FermiSurface}
\sigma_{abc}^{G}
&=\frac{e^2}{\hbar}
\sum_{n}\int_{\bf k}
\bigg[\frac{1}{\omega}(\d_af_{n}B^{bc}_{nn}-\d_bf_{n}B^{ac}_{nn})\notag\\
&-\d_cf_{n}
\sum_m{\rm Im}(r^a_{nm}r^{b}_{mn})\frac{\omega_{mn}\omega}{\omega_{mn}^2-\omega^2}
\bigg].
\end{align}
The effect of dissipation is included by the substitution $\omega\rightarrow \omega+i\tau^{-1}$.

\begin{figure}[t]
\includegraphics[width=0.48\textwidth]{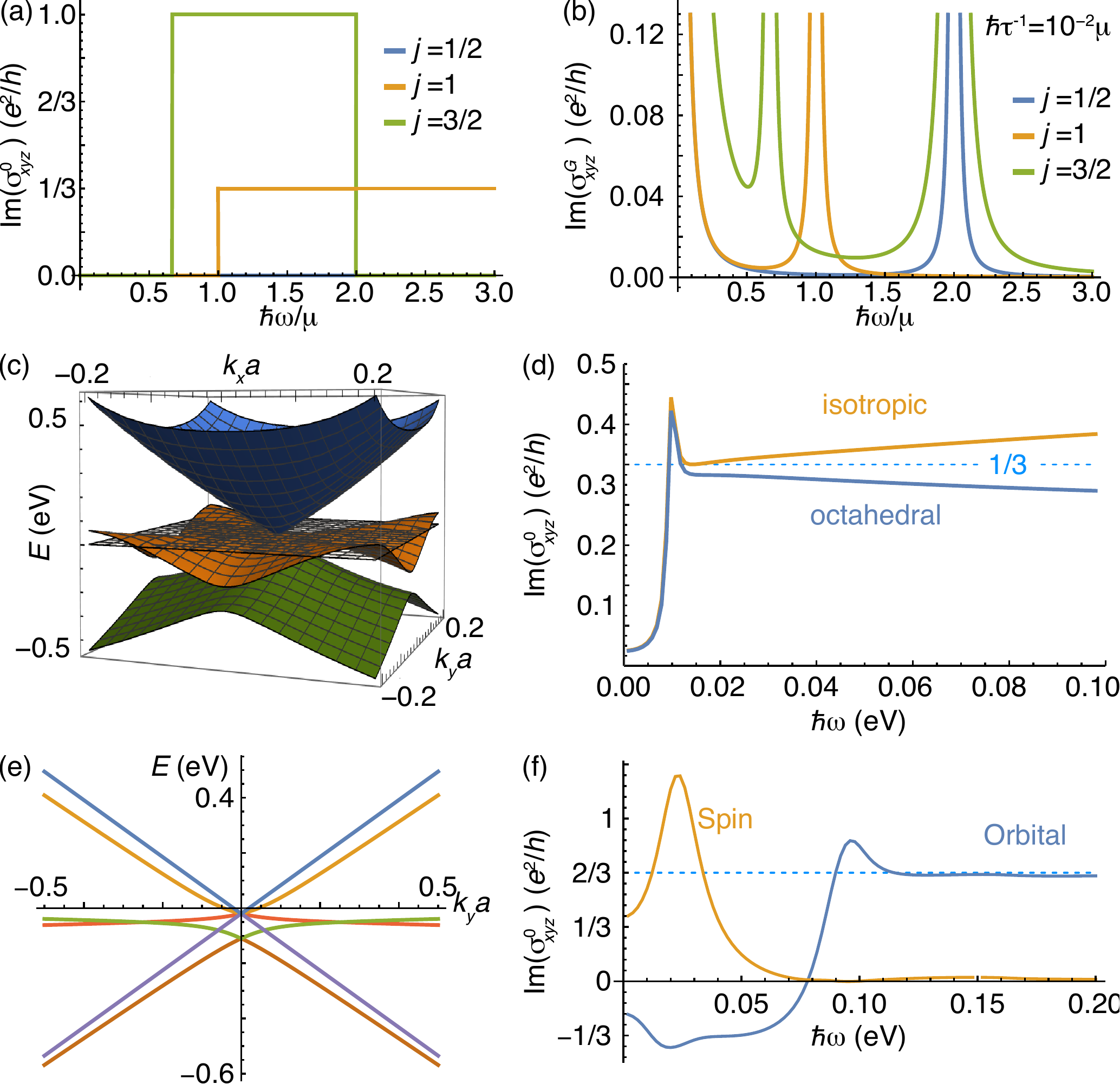}
\caption{
The imaginary part of $\sigma_{xyz}$ of a pseudospin-$j$ fermion.
(a,b) Spinless linearly dispersing fermion.
(a) Fermi sea contribution $\sigma^0_{xyz}$ and
(b) Fermi surface contribution $\sigma^G_{xyz}$ of the linearly dispersing model in Eq.~\eqref{eq:model} without quadratic terms.
While we take $\omega\tau\rightarrow \infty$ in (a), we introduce a finite relaxation time in (b).
We take $\mu>0$ for all plots.
(c,d) Band structure and $\sigma^0_{xyz}$ with quadratic terms in Eq.~\eqref{eq:quadratic}.
$\mu=0.01$ eV, $A=1.07\tilde{a}^2$ eV, $B=-1.72\tilde{a}^2$ eV, $C=3.26\tilde{a}^2$ eV, and $\hbar v=1.79\tilde{a}$ eV, where $\tilde{a}=a_{\rm CoSi}/(2\pi)$, $a_{\rm CoSi}=4.45$ {\AA} is the lattice constant of CoSi, $\hbar\tau^{-1}=1$ meV.
The transparent plane in (c) shows the Fermi level.
The orange curve in (d) shows the isotropic case ($B=C=0$ with other parameters kept unchanged) for comparison.
(e,f) Band structure and $\sigma^0_{xyz}$ with spin-orbit coupling in Eq.~\eqref{eq:spin-orbit coupling}.
$w=30 a$ meV, $\Delta=30$ meV, and $\hbar\tau^{-1}=10$ meV.
In (f), the spin part is due to spin magnetic moment, and the orbital part refers to the other contributions.
}
\label{fig:OA}
\end{figure}

In the clean limit where $\omega\tau\rightarrow \infty$, the Fermi sea part is purely imaginary and thus describes circular dichroism.
\begin{align}
&\sigma^0_{abc}
=\frac{i\pi e^2\omega}{\hbar}
\sum_{\substack{n={\rm occ}\\m={\rm unocc}}}\int_{\bf k}
\delta(\omega_{mn}-\omega)\times
{\rm Im}\notag\\
&\bigg[
(r^a_{nm}B^{bc}_{mn}-r^b_{nm}B^{ac}_{mn}-\frac{1}{2}
(r^{a}_{nm}r^b_{mn})
(v^c_{mm}+v^c_{nn})
\bigg].
\end{align}
For the model in Eq.~\eqref{eq:model}, we obtain quantized values
\begin{align}
\sigma_{abc}^0
&=i\epsilon_{abc}s_{\mu}\chi\frac{e^2}{3h}\notag\\
&\times
\begin{cases}
0                           &j=1/2,\\
\Theta(\hbar\omega-|\mu|)   &j=1,\\
3[\Theta(\hbar\omega-\frac{3|\mu|}{2})-\Theta(\hbar\omega-2|\mu|)]  &j=3/2,
\end{cases}
\end{align}
where $s_{\mu}=\mu/|\mu|$
[Fig.~\ref{fig:OA}(a)].
The Chern number origin of the quantization is manifested in the expression of the nonvanishing value $\sigma^0_{xyz}=-is_{\mu}c_{2j+j}e^2(6h)^{-1}\left|(1+v_{2j}/v_{2j+1})/(1-v_{2j}/v_{2j+1})\right|$, where $c_{2j+1}=-(2\pi)^{-1}\oint d{\bf S}\cdot {\bf F}_{2j+j}=-2\chi j$ is the outward Berry flux of the topmost band, i.e., band $2j+1$~\cite{supp}.
We note that the velocity ratio takes a universal value independent of material specifics, $v_{2j}/v_{2j+1}=-1$, $0$ and $1/3$ for $j=1/2$, $1$, and $3/2$, only when the effective Hamiltonian has isotropy, which requires specific space group symmetries as we discuss above.

The isotropic linearly dispersing Weyl fermion does not show circular dichroism from the Fermi sea~\cite{goswami2015optical}.
This is because of the constraint from $CP$ symmetry that imposes $B^{ab}_{mn}=0$ and $v^a_{mm}+v^a_{nn}=0$ between $CP$-related states $m$ and $n$,
where $C$ is particle-hole conjugation, and $P$ is spatial inversion~\cite{supp}.
The nontrivial circular dichroism of multifold fermions is due to $CP$-asymmetric optical excitations which generate the net change of the orbital magnetic moment.
This favors the absorption of one particular circular polarization of light to the other polarization.

Let us consider shining linearly polarized or unpolarized light.
Then, the incident intensity is the same for $L/R$ helicity on average.
The transmitted light intensity after propagation of the distance $d$ within the material is $I_{L/R}=2^{-1}I_0|\exp(2\pi i n_{L/R}d/\lambda)|^2
$ for $L/R$ helicity, where $I_0$ is the incident light intensity.
The transmissive circular dichroism is defined by
\begin{align}
\label{eq:TCD}
{\rm CD}
\equiv \frac{I_L-I_R}{I_L+I_R}
=\tanh\left(\chi s_{\mu}\frac{4\pi \alpha N_j}{3}\frac{d}{\lambda}\right)
\end{align}
where $\alpha=\mu_0ce^2/2h$ is the fine structure constant.

In the clean limit, the Fermi surface part does not contribute to the circular dichroism because it is real valued, where
\begin{align}
\sigma_{abc}^{G}
=\epsilon_{abc}\chi\frac{e^2}{3\pi h}
\frac{\mu}{\hbar\omega}\left[\frac{\mu^2-3j^2(\hbar\omega)^2/2}{\mu^2-j^2(\hbar\omega)^2}+f_j\right],
\end{align}
and $f_{1/2}=f_1=0$, and $f_{3/2}=7[\mu^2-3(\hbar\omega)^2/8]/[\mu^2-(\hbar\omega)^2/4]$
But this contributes to the circular dichroism when there is a finite relaxation and is proportional to $\tau^{-1}$.
Figure~\ref{fig:OA}(b) shows the case with $\hbar\tau^{-1}=0.01\mu$.

{\it Effect of quadratic dispersion and spin-orbit coupling.---}
To see the effect of $O(k^2)$ terms, we consider $H=H_0+H_{1}$ of a threefold fermion with an additional quadratic Hamiltonian allowed by octahedral symmetry:
\begin{align}
\label{eq:quadratic}
H_{1}
=
\begin{pmatrix}
Xk^2-2Ck_z^2&Bk_yk_z&Bk_zk_x\\
Bk_yk_z&Xk^2-2Ck_x^2&Bk_xk_y\\
Bk_zk_x&Bk_xk_y&Xk^2-2Ck_y^2
\end{pmatrix},
\end{align}
where $X=A+2C/3$, and ${\bf J}=\{-\lambda_2,\lambda_5,-\lambda_7\}$ for $H_0$, and $\lambda_i$ is the Gell-Mann matrix~\cite{ni2021giant}.

Figure~\ref{fig:OA}(c) shows the band structure with quadratic terms included.
We take $\mu=0.1$ eV and the model parameters for CoSi derived in Ref.~\cite{ni2021giant}, which are $A=1.07\tilde{a}^2$ eV, $B=-1.72\tilde{a}^2$ eV, $C=3.26\tilde{a}^2$ eV, and $\hbar v=1.79\tilde{a}$ eV, where $\tilde{a}$ has the dimension of length ($\tilde{a}=a_{\rm CoSi}/2\pi$, where $a_{\rm CoSi}=4.45$ {\AA} is the lattice constant of CoSi).

When the quadratic terms are included, the value of ${\rm Im}(\sigma^0_{xyz})$ deviates from the quantized plateau [Fig.~\ref{fig:OA}(d)].
The deviation originates from the momentum dependence of the velocities of bands~\cite{supp}, and the effect of selection-rule-breaking transitions is negligible (less than 1 \%).
Therefore, an isotropic quadratic dispersion that preserves the selection rule can lead to a comparable deviation from the quantization [orange curve in Fig.~\ref{fig:OA}(d)].

The effect of spin-orbit coupling is twofold. One is the spin-orbit splitting of the band structure, and the other is contribution from the spin magnetic moment.
The former effect is absent in the case where Eq.~\eqref{eq:model} is realized in the presence of spin-orbit coupling.
Here, we consider the case of a threefold fermion realized in each spin sector in the absence of spin-orbit coupling, with application to CoSi in mind.
In this case, the spin-orbit coupling up to linear order in $k$ is given by
\begin{align}
\label{eq:spin-orbit coupling}
H_{\rm SOC}={\bf s}\cdot \left(w{\bf k}+\Delta{\bf J}\right),
\end{align}
where $s_{i=x,y,z}$ is the spin Pauli matrix.
For a threefold (per spin) fermion, this splits the sixfold (including spin) degeneracy into fourfold and twofold degenerate points by $\delta E_{\rm SOC}=3\Delta$ [Fig.~\ref{fig:OA}(e)].
Figure~\ref{fig:OA}(f) shows that the circular dichroism approaches to the quantized value as the photon energy becomes larger than $\delta E_{\rm SOC}$.
The effect of spin magnetic moment is negligible in the quantized regime.

\begin{figure}[t]
\includegraphics[width=0.5\textwidth]{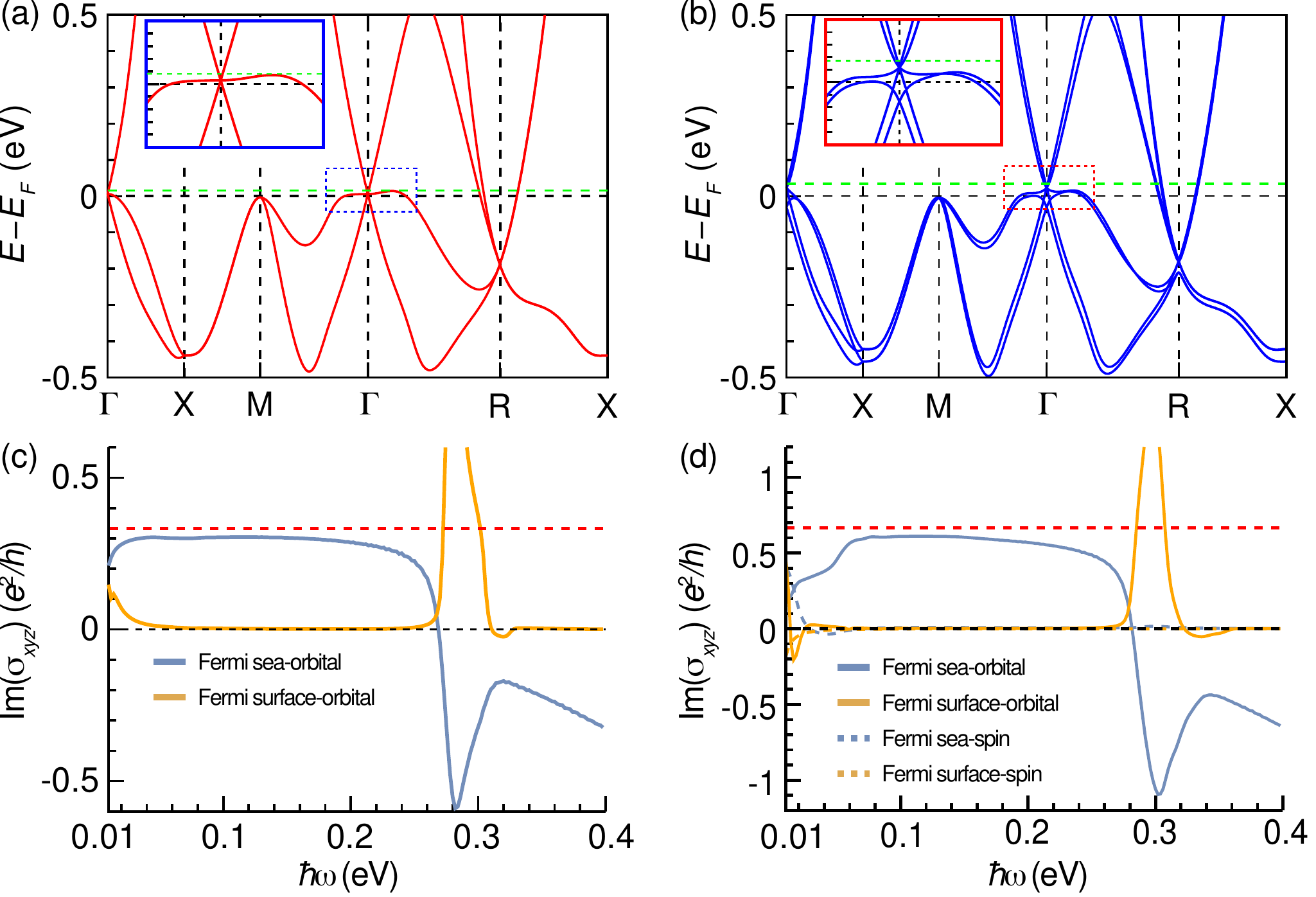}
\caption{Ab-initio calculations for threefold semimetal CoSi based on density functional theory.
(a,b) Band structure. (a) without and (b) with spin-orbit coupling.
The insets show the band structure near the $\Gamma$ point. The horizontal green dashed line denotes the chemical potential used for computing the $\sigma^0_{xyz}$  and $\sigma^G_{xyz}$. 
(c,d) The imaginary parts of Fermi sea ($\sigma^0_{xyz}$) and Fermi-surface ($\sigma^G_{xyz}$) contributions
(c) without and (d) with spin-orbit coupling.
$\hbar\tau^{-1}=1$ meV.
}

\label{fig:DFT}
\end{figure}

{\it Chiral threefold semimetal CoSi.}
We now turn the discussion toward material-specific DFT-based calculations to test our model analysis. We focus on the transition metal monosilicide family of materials CoSi, which crystalizes in the B20 cubic structure~\cite{CoSi_scz,CoSi_arpes_1}. The crystal structure is chiral, and it belongs to the $P2_13$ space group (SG198); it lacks an inversion, mirror, and roto-inversion symmetry. The structural chirality and the octahedral symmetries lead to various types of multifold fermions in these systems~\cite{CoSi_scz,CoSi_arpes_1,CoSi_arpes_2,CoSi_plasmon}. Specifically, in the absence of spin-orbit interaction, CoSi host a threefold degenerate nodal point at the zone center and double Weyl fermion state at the corner of the cubic BZ [Fig.~\ref{fig:DFT}(a)].

We compute ${\rm Im}(\sigma^0_{xyz})$ for CoSi using the Wannier function-based tight-binding model [see Supplemental Material for details]. The chemical potential (indicated by the green dashed line) is set to be slightly above the threefold degenerate crossing point to ensure full occupancy of the flat band around the $\Gamma$-point. The tuning of the chemical potential has been experimentally achieved recently in RhSi via Ni doping ~\cite{RhSi_doping}. As shown in Fig.~\ref{fig:DFT}(c), the calculated ${\rm Im}(\sigma^0_{xyz})$ results strongly support our low energy model analysis. Specifically, we found that in CoSi, the ${\rm Im}(\sigma^0_{xyz})$ starts from a finite value for low photon energy and it quickly approaches  the quantized value $e^2/3h$, developing a plateau-like region for $50\lesssim\hbar \omega\lesssim 200$ meV. In this region, the optical transitions involving the threefold fermion around the $\Gamma$ point plays the important role. The small deviation from the quantized value is attributed to the presence of quadratic band dispersion, and it supports our model analysis. For $\hbar \omega\gtrsim 200$ meV,  the optical transitions involving the states around the R point become important, and consequently, the ${\rm Im}(\sigma^0_{xyz})$ changes sign, as it strongly deviates from the quantized value.  For comparison, we also compute the Fermi surface contribution ${\rm Im}(\sigma^G_{xyz}$) for CoSi, which was studied in a previous work~\cite{flicker2018chiral}. In general, ${\rm Im}(\sigma^G_{xyz}$) is smaller compared to the ${\rm Im}(\sigma^0_{xyz}$), and its value depends strongly on the relaxation time, and in the clean limit $\omega\tau \gg 1$. This Fermi surface contribution should be negligible in the quantized region.

We further consider the effect of spin-orbit coupling in Fig.~\ref{fig:DFT}(b,d).
In consistent with model analysis, approximate quantization of ${\rm Im}(\sigma^0_{xyz})$ still holds true even after including the effect of spin-orbit coupling, and
the spin magnetic moment contributes negligibly compared to the orbital part in the plateau region.

We also explored other material candidates in this family, including RhSi, and PtAl (see Supplemental Material~\cite{PtAl_arpes_2,RhSn_expt}). Our analysis suggests that in the absence of spin-orbit coupling, the approximate quantization of ${\rm Im}(\sigma^0_{xyz})$ holds true both in RhSi and PtAl. However, due to the presence of large spin-orbit coupling in these compounds, the ${\rm Im}(\sigma^0_{xyz})$ deviates from the quantized value. Interestingly, this deviation is still approximately within 10 \% for RhSi and 20 \% for PtAl, despite the spin-orbit coupling being significantly stronger compared to CoSi.

{\it Conclusion.---}
Our analysis establishes that topological circular dichroism is the unique feature of multifold fermions in the k dot p regime.
Thin films will be ideal for an observation of this effect because transmitted light intensity is exponentially suppressed in bulk samples.
Topological circular dichroism is similar to the quantized absorption in graphene~\cite{nair2008fine} because it requires linear dispersion.
The quantization is expected to be robust as long as photon energy is much larger than thermal energy. However, disorder and interaction effects can give deviations from quantized optical responses~\cite{avdoshkin2020interactions,mandal2020effect}, in contrast to the quantum Hall effect.
We leave detailed analysis of these effects for future studies.


\begin{acknowledgments}
We appreciate Ashvin Vishwanath, Arun Bansil, Su-Yang Xu, and Yuan Ping for helpful discussions.
We thank Ipsita Mandal for bringing their work~\cite{sekh2022circular,mandal2023signatures} to our attention when our manuscript is being finalized.
J.A. was supported by the Center for Advancement of Topological Semimetals, an Energy Frontier Research Center funded by the U.S. Department of Energy Office of Science, Office of Basic Energy Sciences, through the Ames Laboratory under contract No. DE-AC02-07CH11358.
B.G. was supported by the Air Force Office
of Scientific Research under Award No. FA9550-20-1-0322 and benefited from the computational resources of Northeastern University’s Advanced Scientific Computation Center
(ASCC) and the Discovery Cluster.
\end{acknowledgments}


%

\clearpage
\newpage

\renewcommand{\thefigure}{S\arabic{figure}}
\renewcommand{\theequation}{S\arabic{equation}}
\renewcommand{\thetable}{S\arabic{table}}
\renewcommand{\thesubsection}{Supplementary Note \arabic{subsection}}

\setcounter{figure}{0} 
\setcounter{equation}{0} 
\setcounter{table}{0} 
\setcounter{section}{0} 

\begin{widetext}

\section{Conductivity formula}

The bulk current density is related to the external electric field by
\begin{align}
J_a(\omega,{\bf q})
&=\sigma_{ab}(\omega,{\bf q})E_b(\omega,{\bf q})
\end{align}
in linear response theory.
The conductivity is given by the Kubo-Greenwood formula~\cite{malashevich2010band}
\begin{align}
\sigma_{ab}(\omega,{\bf q})
&=
-\frac{ie^2}{\hbar \omega}
\sum_{n,m}\int_{\bf k}\frac{f_{n({\bf k}-{\bf q}/2)}-f_{m({\bf k}+{\bf q}/2)}}{\omega_{m({\bf k}+{\bf q}/2)}-\omega_{n({\bf k}-{\bf q}/2)}-\omega}
\braket{\psi_{n({\bf k}-{\bf q}/2)}|\hat{I}_a^{\dagger}({\bf q})|\psi_{m({\bf k}+{\bf q}/2)}}
\braket{\psi_{m({\bf k}+{\bf q}/2)}|\hat{I}_b({\bf q})|\psi_{n({\bf k}-{\bf q}/2)}}\notag\\
&=\sigma_{ab}(\omega)+\sigma_{abc}(\omega)q_c+O(q^2).
\end{align}
The $q$-linear expansion of the Fermi-Dirac distribution $f_n$ is due to the Fermi surface.
\begin{align}
\sigma^{(1)}_{abc}
&=-\frac{ie^2}{\hbar \omega^2}
\sum_{n}\int_{\bf k}
\d_cf_{n}
v^{a}_{nn}v^b_{nn}
+\frac{ie^2}{\hbar}
\sum_{n,m}\int_{\bf k}
\d_cf_{n}
{\rm Re}[r^a_{nm}r^{b}_{mn}]\frac{\omega_{mn}^2}{\omega_{mn}^2-\omega^2}\notag\\
&\qquad-\frac{e^2}{\hbar \omega}
\sum_{n,m}\int_{\bf k}
\d_cf_{n}
\omega_{mn}{\rm Im}(r^a_{nm}r^{b}_{mn})\frac{\omega_{mn}^2}{\omega_{mn}^2-\omega^2}.
\end{align}
In insulators, only the other terms contribute to the response.
\begin{align}
\sigma^{(2)}_{abc}
&=-\frac{ie^2}{\hbar}
\sum_{n,m}\int_{\bf k}
f_{nm}
\left[
\frac{\omega_{mn}}{\omega(\omega_{mn}-\omega)}
\left((r^b_{nm}B^{ac}_{mn})^*
+r^a_{nm}B^{bc}_{mn}\right)
-\frac{1}{\omega}\frac{\omega_{mn}^2}{2(\omega_{mn}-\omega)^2}
r^{a}_{nm}r^b_{mn}(v^c_{mm}+v^c_{nn})
\right],
\end{align}
where we use that
\begin{align}
\braket{\psi_{m({\bf k}+{\bf q}/2)}|\hat{I}_a({\bf q})|\psi_{n({\bf k}-{\bf q}/2)}}
&=v^a_{mn}({\bf k})+iq_cB^{ac}_{mn}+O(q^2),
\end{align}
and
\begin{align}
B^{ac}_{mn}
=\frac{1}{2}\left(\braket{iD_cu_{m{\bf k}}|\hat{v}^a_{\bf k}|u_{n{\bf k}}}
+\braket{u_{m{\bf k}}|\hat{v}^a_{\bf k}|iD_cu_{n{\bf k}}}\right)
-\frac{g_s}{2m_e}\epsilon_{ikl}\braket{u_{m{\bf k}}|\hat{S}^l|u_{n{\bf k}}}
=iB^{{\rm MS},ik}_{mn},
\end{align}
where $\ket{D_cu_{n{\bf k}}}=\ket{\partial_cu_{n{\bf k}}}-i\sum_{m}\ket{u_{m{\bf k}}}A_{mn}({\bf k})$, and ${\bf A}_{mn}({\bf k})=\delta_{E_m,E_n}\braket{u_{m{\bf k}}|i\nabla_{\bf k}|u_{m{\bf k}}}$ is the Berry connection.
We define $B^{ac}$ as a Hermitian matrix in accordance with the notation of Pozo and Souza~\cite{pozo2023multipole} while the same notation was originally used to define an anti-Hermitian matrix by Malashevich and Souza (MS)~\cite{malashevich2010band}, which we write as $B^{{\rm MS},ac}$.

We define symmetrized and antisymmetrized tensors by
\begin{align}
\sigma_{abc}^S
&=\frac{\sigma_{abc}+\sigma_{jik}}{2},\notag\\
\sigma_{abc}^A
&=\frac{\sigma_{abc}-\sigma_{jik}}{2}.
\end{align}
It follows that
\begin{align}
\sigma_{abc}^S
&=-\frac{ie^2}{\hbar}
\sum_{n,m}\int_{\bf k}
f_{nm}
\left[
\frac{\omega_{mn}}{\omega_{mn}^2-\omega^2}
{\rm Re}\left[r^b_{nm}B^{ac}_{mn}
+r^a_{nm}B^{bc}_{mn}\right]
-\frac{\omega_{mn}^3}{(\omega_{mn}^2-\omega^2)^2}
{\rm Re}\left[r^{a}_{nm}r^b_{mn}\right](v^c_{mm}+v^c_{nn})
\right],\notag\\
&\qquad-\frac{ie^2}{\hbar \omega^2}
\sum_{n,m}\int_{\bf k}
\d_cf_{n}
\left(v^{a}_{nn}v^b_{nn}
-\omega^2{\rm Re}[r^a_{nm}r^{b}_{mn}]\frac{\omega_{mn}^2}{\omega_{mn}^2-\omega^2}
\right),
\end{align}
where the last term describes the quantum metric contribution to the nonreciprocal directional dichroism~\cite{gao2019nonreciprocal,lapa2019semiclassical}, and
\begin{align}
\label{eq:sigma-A}
\sigma_{abc}^A
&=\sigma_{abc}^0+\sigma_{abc}^G\notag\\
&=\frac{e^2}{\hbar}
\sum_{n,m}\int_{\bf k}
f_{nm}
\left[
\frac{\omega}{\omega_{mn}^2-\omega^2}
{\rm Im}\left[-r^b_{nm}B^{ac}_{mn}
+r^a_{nm}B^{bc}_{mn}\right]
-\frac{(3\omega_{mn}^2-\omega^2)\omega}{2(\omega_{mn}^2-\omega^2)^2}
{\rm Im}\left[r^{a}_{nm}r^b_{mn}\right](v^c_{mm}+v^c_{nn})
\right]\notag\\
&\qquad+\frac{e^2}{\hbar}
\sum_{n,m}\int_{\bf k}
\left(\frac{1}{\omega}(-\d_bf_{n}B^{ac}_{nn}+\d_af_{n}B^{bc}_{nn})-\d_cf_{n}
{\rm Im}(r^a_{nm}r^{b}_{mn})\frac{\omega\omega_{mn}}{\omega_{mn}^2-\omega^2}
\right),
\end{align}
where $\sigma_{abc}^0$ is the Fermi sea contribution, and $\sigma_{abc}^G$ is the Fermi level contribution responsible for the gyrotropic magnetic effect~\cite{ma2015chiral,zhong2016gyrotropic}.
The gyrotropic magnetic effect refers to to the generation of electric current by slowly time-varying magnetic field through
\begin{align}
J_a(\omega)=\alpha_{ab}(\omega)B_b(\omega).
\end{align}
The small-frequency limit (still in the clean limit $\omega\tau\gg 1$) of $\alpha(\omega)$ is given by
\begin{align}
\alpha_{ab}(\omega\ll \omega_{mn})
&=\lim_{\omega/\omega_{mn}\rightarrow 0}\left[-\frac{\omega}{4}\sum_{c,d}\epsilon_{bcd}(2\sigma_{acd}-\sigma_{cda})\right]\notag\\
&=-\frac{e}{\hbar}
\sum_{n}\int_{\bf k}\d_af_{n}\left(-\frac{e}{2}\sum_{c,d}\epsilon_{bcd}B^{cd}\right)\notag\\
&=-\frac{e}{\hbar}
\sum_{n}\int_{\bf k}\d_af_{n}m^b_n,
\end{align}
where
$m^{b}_n=m^{{\rm orb},b}_n+m^{{\rm spin},b}_n=\sum_{k,l}(e/2)\epsilon_{bcd}\sum_m\omega_{mn}{\rm Im}(r^c_{nm}r^{d}_{mn})+(-eg_s/2m_e)S_n^b=
(e/2)\sum_{c,d}\epsilon_{bcd}B^{cd}$ is the magnetic moment of band $n$.

\section{$CP$ symmetry constraints on multipole transitions}

Here we generalize the $CP$ selection rule for electric dipole transitions in Ref.~\cite{ahn2021theory,ahn2021many} to general electromagnetic multipole transitions.
Let $\hat{O}$ be a Hermitian operator.
\begin{align}
\braket{CP \psi_{n{\bf k}}|\hat{O}({\bf k})\psi_{n{\bf k}}}
&=\braket{CP\hat{O}({\bf k})\cdot \psi_{n{\bf k}}|(CP)^2 \psi_{n{\bf k}}}\notag\\
&=(CP)^2\braket{CP \psi_{n{\bf k}}|CP\hat{O}({\bf k})(CP)^{-1}|\psi_{n{\bf k}}}\notag\\
&=\eta_{O}\epsilon_{CP}\braket{CP \psi_{n{\bf k}}|\hat{O}({\bf k})|\psi_{n{\bf k}}}.
\end{align}
We use that $CP$ is a anti-unitary operator in the first line, use that $(CP)^2=\epsilon_{CP}=\pm 1$ is a number, and $\hat{O}$ is Hermitian in the second line, and define $\eta_{O}=\pm 1$ by $CP\hat{O}({\bf k})CP^{-1}=\eta_{O}\hat{O}({\bf k})$ in the third line.
$\eta_O=(-1)^{N+1}$ when $\hat{O}$ is the $N$th-order electric multipole ($EN$) moment, and $\eta_O=(-1)^N$ when $\hat{O}$ is the $N$th-order magnetic multipole ($MN$) moment, where the first-order multipole is the dipole.

The absence of interband natural optical activity of an ideal Weyl fermion is due to this constraint.
The two bands of a linearly dispersing Weyl fermion are related by $CP$ with respect to the nodal point because the Hamiltonian has $CP$ symmetry $CPH({\bf k})(CP)^{-1}=-H({\bf k})$ at $\mu=0$, where $CP=i\sigma_yK$, and $\sigma_y$ is the pseudospin Pauli matrix.
Because $(CP)^2=-1$, we have $\braket{CP\psi_{n\bf k}|\hat{M}_a|u_{n\bf k}}=\braket{CP\psi_{n\bf k}|\hat{Q}_{ab}|\psi_{n\bf k}}=0$ .
Therefore, $\sigma_{abc}=0$ by this constraint because $\sigma_{abc}$ is due to the combination of electric dipole and magnetic dipole ($E1$-$M1$) or the combination of electric dipole and electric quadrupole ($E1$-$E2$) transitions~\cite{malashevich2010band}.

\section{Model calculations}

Let us consider the spin-1 Hamiltonian.
\begin{align}
\label{eq:model}
H=-\mu({\bf k})+
i\hbar \chi v\begin{pmatrix}
0&k_x&-k_y\\
-k_x&0&k_z\\
k_y&-k_z&0
\end{pmatrix},
\end{align}
where $\chi=\pm 1$ determines the chirality.
The energy eigenvalues are
\begin{align}
E_1&=-\mu-\chi\hbar vk,\notag\\
E_2&=-\mu,\notag\\
E_3&=-\mu+\chi\hbar vk,
\end{align}
and the corresponding eigenstates are
\begin{align}
\ket{1}
&=\frac{1}{\sqrt{2}k\sqrt{k_x^2+k_z^2}}
\begin{pmatrix}
k_yk_z+ikk_x\\
-k_x^2-k_z^2\\
k_xk_y-ikk_z
\end{pmatrix},\notag\\
\ket{2}
&=\frac{1}{k}
\begin{pmatrix}
k_z\\
k_y\\
k_x
\end{pmatrix},\notag\\
\ket{3}
&=\frac{1}{\sqrt{2}k\sqrt{k_x^2+k_z^2}}
\begin{pmatrix}
k_yk_z-ikk_x\\
-k_x^2-k_z^2\\
k_xk_y+ikk_z
\end{pmatrix}.
\end{align}
Suppose that $\mu>0$ and $\chi=1$.
Then, $\sigma^0_{xyz}$ for the chiral threefold fermion is
\begin{align}
\sigma_{xyz}^0
&=-\frac{2e^2}{\hbar}
\int_{\bf k}
\left[
-\frac{\omega}{\omega_{32}^2-\omega^2}
2{\rm Im}\left[r^x_{23}B^{yz}_{32}\right]
+\frac{(3\omega_{32}^2-\omega^2)\omega}{2(\omega_{32}^2-\omega^2)^2}
{\rm Im}\left[r^{x}_{23}r^y_{32}\right](v^z_{3}+v^z_{2})
\right]\notag\\
&=-\frac{2e^2}{\hbar}
\int_{\bf k}
\left[
-\frac{\omega}{\omega_{32}^2-\omega^2}
{\rm Im}\left[r^x_{23}r^{z}_{32}\right](v^y_{3}+v^y_{2})
+\frac{(3\omega_{32}^2-\omega^2)\omega}{2(\omega_{32}^2-\omega^2)^2}
{\rm Im}\left[r^{x}_{23}r^y_{32}\right](v^z_{3}+v^z_{2})
\right]\notag\\
&=-\frac{2e^2}{\hbar}
\int_{\bf k}
\frac{(5\omega_{32}^2-3\omega^2)\omega}{2(\omega_{32}^2-\omega^2)^2}
{\rm Im}\left[r^x_{23}r^{y}_{32}\right](v^z_{3}+v^z_{2})\notag\\
&=\frac{2e^2}{\hbar}
v\omega\int \frac{d\phi d\theta  \sin\theta dkk^2}{(2\pi)^3}
\frac{5\omega_{32}^2-3\omega^2}{2(\omega_{32}^2-\omega^2)^2}
\frac{1}{2k^2}\cos^2\theta\notag\\
&=\frac{e^2v\omega}{4\pi^2\hbar}
\int^{\pi}_0 d\theta  \sin\theta\cos^2\theta \int^{\infty}_0 dk
\frac{5\omega_{32}^2-3\omega^2}{2(\omega_{32}^2-\omega^2)^2}\notag\\
&=\frac{ie^2}{6\pi\hbar}
\end{align}
in the clean limit, where we use $ {\rm Im}(r^x_{23}r^y_{32})=-k_z/2k^3$, $v^z_3=k_z/k$, $v^z_2=0$, and $\int^{\infty}_0 dx (5x^2-3\omega^2)/(x^2-\omega^2)^2/2=i\pi/\omega$ for ${\rm Im}(\omega)>0$.
We get an opposite sign for $\mu<0$.

We can rewrite the above derivation to see which properties of the model we use in the derivation.
\begin{align}
\sigma^0_{xyz}
&=-s_{\mu}\frac{e^2}{\hbar}
\int_{\bf k}
\frac{(5\omega_{32}^2-3\omega^2)\omega}{2(\omega_{32}^2-\omega^2)^2}
F^z_{3}(v^z_{3}+v^z_{2})\notag\\
&=-s_{\mu}\frac{e^2}{3\hbar}
\int_{\bf k}
\frac{(5\omega_{32}^2-3\omega^2)\omega}{2(\omega_{32}^2-\omega^2)^2}
{\bf F}_3\cdot\hat{n}|\d_{\bf k}(\omega_c+\omega_v)|
\notag\\
&=-s_{\mu}\frac{e^2\omega}{3(2\pi)^2\hbar}
\frac{1}{2\pi}\int_{k_\perp} d{\bf S}\cdot {\bf F}_3\int dk_\perp |\d_{\bf k}(\omega_c+\omega_v)|
\frac{5\omega_{32}^2-3\omega^2}{2(\omega_{32}^2-\omega^2)^2}
\notag\\
&=-s_{\mu}c_{3}\frac{e^2\omega}{3(2\pi)^2\hbar}
\int d\omega_{32}
\left(\frac{v_3^{\perp}+v_2^{\perp}}{v_3^{\perp}-v_2^{\perp}}\right)
\frac{5\omega_{32}^2-3\omega^2}{2(\omega_{32}^2-\omega^2)^2}\notag\\
&=-\frac{ie^2}{12\pi\hbar}s_{\mu}c_{3}\left(\frac{v_3+v_2}{v_3-v_2}\right).
\end{align}
where $s_{\mu}=\mu/|\mu|$, and $F^a_3=-\sum_{b,c}\epsilon_{abc}{\rm Im}(r^b_{32}r^{c}_{23})$ is the Berry curvature of band $3$, $c_{3}$ is the outward Berry flux of band $3$.
We use $\sigma_{xyz}=\sigma_{yzx}=\sigma_{zxy}$ in the second line, use isotropy of the spectrum in the fourth line, and use the linearity of the spectrum (velocities are constant in momentum space) in the last line.
When quadratic terms are included in the Hamiltonian,
the non-quantization and frequency dependence appears mostly because the velocity ratio $\frac{v_3^{\perp}+v_2^{\perp}}{v_3^{\perp}-v_2^{\perp}}$ is not constant in momentum space any more.

By doing the same calculation for $j=1/2$ and $j=3/2$, we obtain the following expression
\begin{align}
\sigma^0_{xyz}
&=-\frac{ie^2}{12\pi\hbar}s_{\mu}c_{2j+1}\left(\frac{v_{2j+1}+v_{2j}}{v_{2j+1}-v_{2j}}\right).
\end{align}
holds for linearly dispersing pseudospinspin-$j$ fermions with $j=1/2$, $1$, and $3/2$, where $c_{2j+1}=-2\chi j$, and $\chi=\pm 1$ is the chirality of the fermion.

The Fermi-surface part is given by
\begin{align}
\sigma_{abc}^{G}
=\epsilon_{abc}\chi\frac{e^2}{3\pi h}
\frac{\mu}{\hbar\omega}\left[\frac{\mu^2-3j^2(\hbar\omega)^2/2}{\mu^2-j^2(\hbar\omega)^2}+f_j\right]
\end{align}
where $f_{1/2}=f_1=0$, and $f_{3/2}=7[\mu^2-3(\hbar\omega)^2/8]/[\mu^2-(\hbar\omega)^2/4]$

\section{Circular-dichroic photogalvanics}

\begin{figure}[t]
\includegraphics[width=\textwidth]{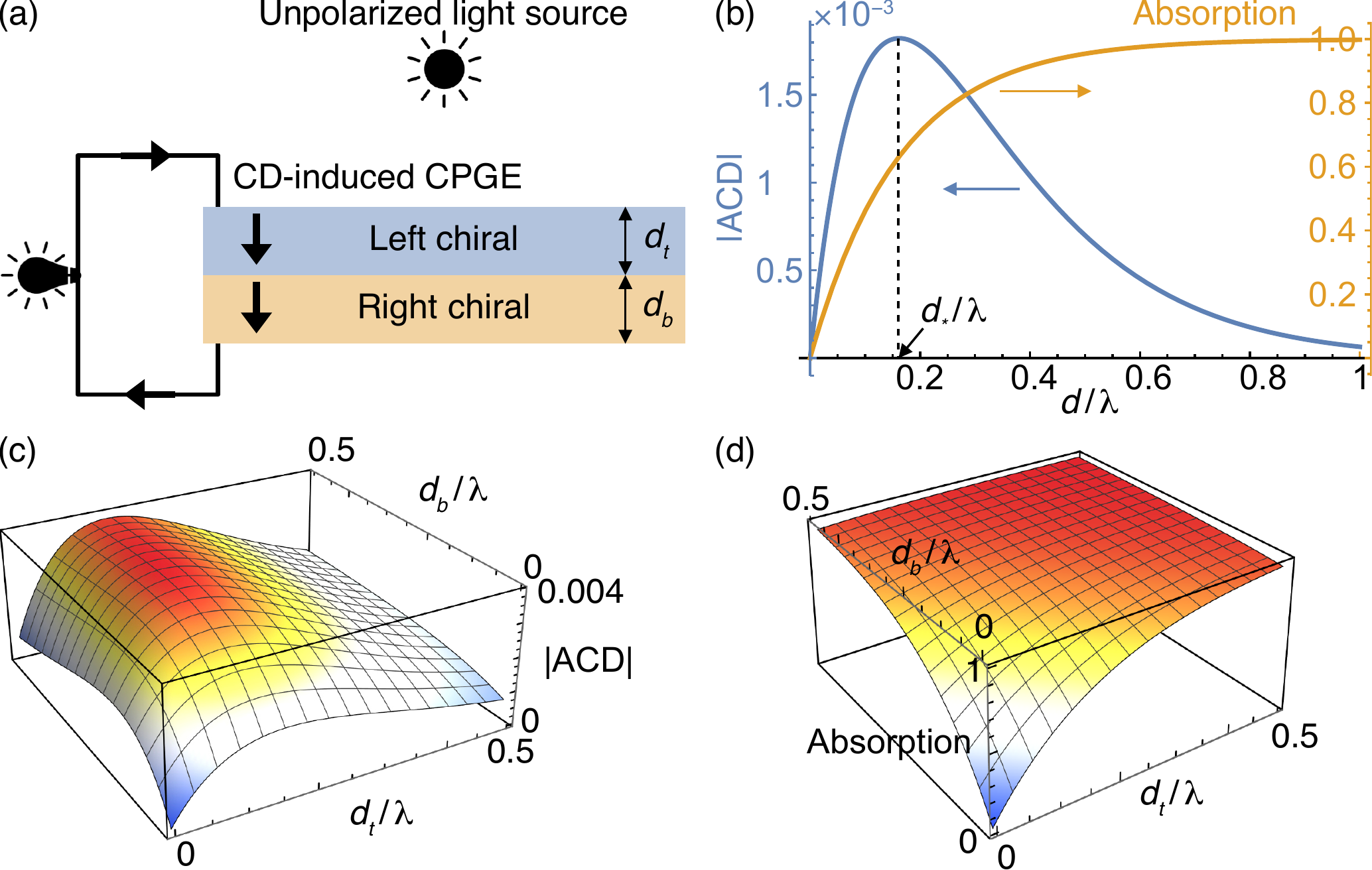}
\caption{
Circular-dichroic photogalvanic effect.
(a) Proposed device geometry.
The first layer, which is left-chiral, absorbs left circularly polarized light more, generating a net DC photocurrent through a circular photogalvanic effect.
The remaining circular polarized light generates the DC photocurrent at the second layer, which is right chiral.
(b) Thickness dependence of absorption for a triple-point fermion semimetal.
The blue and orange curves show the absorptive circular dichroism and total absorption, respectively.
The absolute value of the ACD peaks at $d=d_*$.
(c) Absorptive circular dichroism of the heterochiral bilayer.
(d) Total absorption of the heterochiral bilayer.
We use our model in Eq.~\eqref{eq:model} with $v=0.01c$, where $c$ is the speed of light.
$\hbar\omega>|\mu|$ is assumed in (b-d).
We take $\chi_t>0$, $\chi_b<0$, and $\mu_t \mu_b>0$, and $\omega\tau\rightarrow \infty$.
}
\label{fig:CD-CPGE}
\end{figure}

The imbalance in absorption for different helicities implies that it is possible to generate the circular photogalvanic effect without net helicity of incident light.
While the circular photogalvanic effect leads to much larger photocurrent than the linear photogalvanic effect, it cannot be directly used for photodetection or energy conversion for linearly polarized or unpolarized light, because the photocurrents from two oppositely helical polarizations cancel each other.
However, since chiral materials absorb light with one helicity more than the other, circular dichroism allows the circular photogalvanic effect with incident light having compensated helicity.
Although the circular dichroism is a small effect of about 0.1 \%, the resulting photogalvanic effect can be non-negligible because the circular photogalvanic effect is much larger than the linear photogalvanic effect when the relaxation time is long, by the factor of $\omega\tau$.

We consider the device with two heterochiral materials in Fig.~\ref{fig:CD-CPGE}(a).
A key point here is that the chiral materials should not be too thick.
Let us first consider a single chiral crystal whose low-energy effective model is Eq.~\eqref{eq:model}.
Figure~\ref{fig:CD-CPGE}(a) shows the absorptive dichroism (ACD) for $\chi>0$ (left-chiral) and $\mu<0$, where we define
${\rm ACD}\equiv (A_L-A_R)/I_0$, and $A_{L/R}=I_0-I^t_{L/R}$.
${\rm ACD}>0$ in our case.
Here, $I_0$ is the incident light intensity minus the light intensity reflected at the top.
We neglect the reflection at the bottom for the moment.
The ${\rm ACD}$ is maximal when the thickness of the sample is $d=d_*=(\lambda/8\pi n^i_1)\log(n^i_{L}/n^i_{R})=\lambda/4\pi n^i_0+O[(n^i_1)^2]$, where $n_{L/R}=n_0\pm n_1$ and $n^i_a={\rm Im}(n_a)$, and it approaches zero in thick bulk samples because they perfectly absorbs both helical lights.
Using $\chi_{xx}
=i\alpha(3c/2v)\Theta(\hbar\omega-\mu)$, we have
$d_*=(2\pi)^{-1}(v/3c\alpha)^{1/2}\lambda$, and $|{\rm ACD}|_{d=d_*}=e_0^{-1}(n^i_1/n^i_0)$ in the leading order in $n_1$, where $e_0=2.718\dots$ is the Euler's number.

At the optical thickness $d_*(\lambda)$, only about $1-e_0^{-1}+O(n_1^2)=0.632$ portion of $I_0$ is absorbed.
We can thus improve the device efficiency by adding another chiral material to exploit the transmitted light.
Since the intensity of the right-helical (or left-helical if ${\rm ACD}<0$) light is stronger in transmission, we can put a second chiral material that absorbs right-helical light preferentially.
For example, we take a chiral crystal described by Eq.~\eqref{eq:model} with $\chi<0$ (right-chiral), and $\mu<0$.
To minimize reflections between the top ($t$) and bottom ($b$) chiral crystals, we take $v_t=v_b$ such that the refractive indices are almost identical.
Figure~\ref{fig:CD-CPGE}(c) shows the dependence of the absolute value of ${\rm ACD}$ on $d_t$ and $d_b$.
It peaks at $d_t=0.1364$, and $d_b=0.2985$ in our model, where 93 \% of $I_0$ is absorbed.

While the conditions for maximal ${\rm ACD}$ and maximal circular photogalvanic current coincides in the above example, they are different in general.
The circular photogalvanic current is generated because the group velocity of electronic quasiparticles changes during the optical excitation.
Therefore, the optimization of the circular-dichroism-induced photogalvanic effect depends on the average velocity change as well as the circular dichroism.
This complication goes away as we take $v_t=v_b$ above for a simple demonstration. 

\section{Details of the DFT and Wannier function-based calculations and supplemental figures}

The density functional theory (DFT) based first-principles calculations were performed using the projector augmented wave (PAW) pseudopotentials as implemented in the VASP package~\cite{vasp1,vasp2}. The kinetic energy cutoff for the plane wave basis was set to 400 eV. The generalized gradient approximation (GGA) scheme developed by Perdew-Burke-Ernzerhof (PBE) was used to treat the exchange-correlation part of the potential~\cite{gga_simple}. Experimental lattice parameters were used. A $12\times12\times12$ $\Gamma$-centred $k$-mesh was used for the Brillouin zone integration.

The Wannier function-based tight-binding parametrization was done by using the Wannier90 code~\cite{wannier90}. For computing the Wannier models of CoSi and RhSi, we used projections on Co/Rh $d$ and Si $p$ orbitals, while for PtAl, we included Pt $d$ and Al $s$ and $p$ orbitals. The computation of $\sigma^0_{xyz}$ and $\sigma^G_{xyz}$ was carried out using 11025602 ${\bf k}$-points in the irreducible BZ, which is equivalent to a $640\times640\times640$ ${\bf k}$-mesh in the full BZ. The broadening parameter $\hbar\tau^{-1}$ was set to 6 meV for the calculations, unless specified.

Our method of computing $\sigma^0$ and $\sigma^G$ for realistic materials is based on the Wannier function-based tight binding scheme~\cite{wannier_rmp}. Within this framework, a set of $N$ Bloch states can be described by using $N$ Wannier functions per unit cell as an effective basis. We denote the Bloch basis and the Wannier basis by the superscript ``(H)", and ``(W)", respectively. Following the convention of Ref~\cite{shg_wann_souza,shg_wann_2,Te_nac}, the ${\bf k}$-dependent Hamiltonian in the Wannier basis can be obtained as,
\begin{align}
H^{(W)}_{mn}=\sum_R e^{i{\bf k}\cdot({\bf R+\bm \tau}_n-{\bm \tau}_m)} \langle {\bf 0}m|\hat{H}|{\bf R}n \rangle,
\end{align}
where ${\bm \tau}_n=\braket{{\bf 0}n|\hat{\bf r}|{\bf 0}n}$ is the Wannier center (not the relaxation time $\tau$) of the Wannier state $\ket{{\bf 0}n}$.
In the Bloch basis, $H^{(H)}$ is diagonal and it is  given by,
\begin{align}
 H^{(H)}=U^\dag H^{(W)} U.
\end{align}
The interband $(n\neq m)$ position matrix in the Bloch basis can be calculated using,
\begin{align}
r^{a(H)}=U^\dag r^{i(W)} U + i U^\dag \partial_a U,
\label{rh}
\end{align}
where the position matrix element in the Wannier basis is given by,
\begin{align}
r^{a(W)}_{mn}=\sum_{\bf R} e^{i{\bf k}\cdot({\bf R}+{\bm \tau}_n-{\bm \tau}_m)} \langle {\bf 0}m|\hat{r}_a-\tau_n|{\bf R}n\rangle.
\label{rw}
\end{align} 
It should be noted that the first term in Eq.~\eqref{rh} vanishes $(r^{a(W)}=0)$ if the position operator is diagonal in the Wannier basis, in the so-called ``diagonal tight binding approximation". In our computation, we included the effect of this non-zero off-diagonal position matrix element. 
The second term of Eq.~\eqref{rh} is computed using (for $n\neq m)$,
\begin{align}
(U^\dag \partial_aU)_{nm}=-\frac{(U^\dag v_a^{(W)} U)_{nm}}{E_n-E_m}.
\end{align}
Here the $v^{(W)}_a$ is the velocity matrix in the Wannier basis. 
The interband velocity matrix in the Bloch basis is calculated using,
\begin{align}
v^{a(H)}_{mn}=i\omega_{mn} r^{a(H)}_{mn}
\end{align}
Using these position and velocity matrix elements, we compute the $\sigma^0$ and $\sigma^G$ using Eqs.~(4) and~(5), respectively.

\begin{figure}[ht!]
\includegraphics[width=0.5\textwidth]{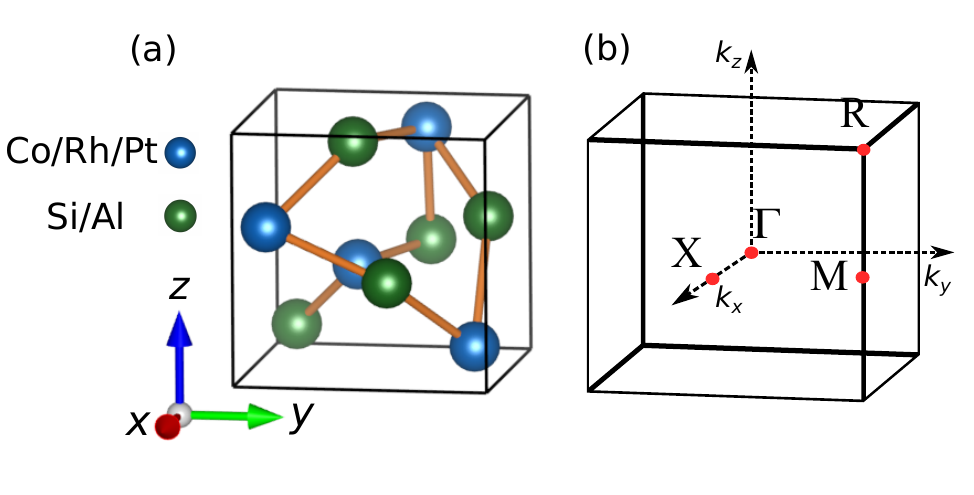}
\caption{Crystal structure and Brillouin zone of CoSi/RhSi/PtAl. In panel (b) the relevant high symmetry points are marked in red.}
\label{fig:CS}
\end{figure}

\begin{figure}[ht!]
\includegraphics[width=\textwidth]{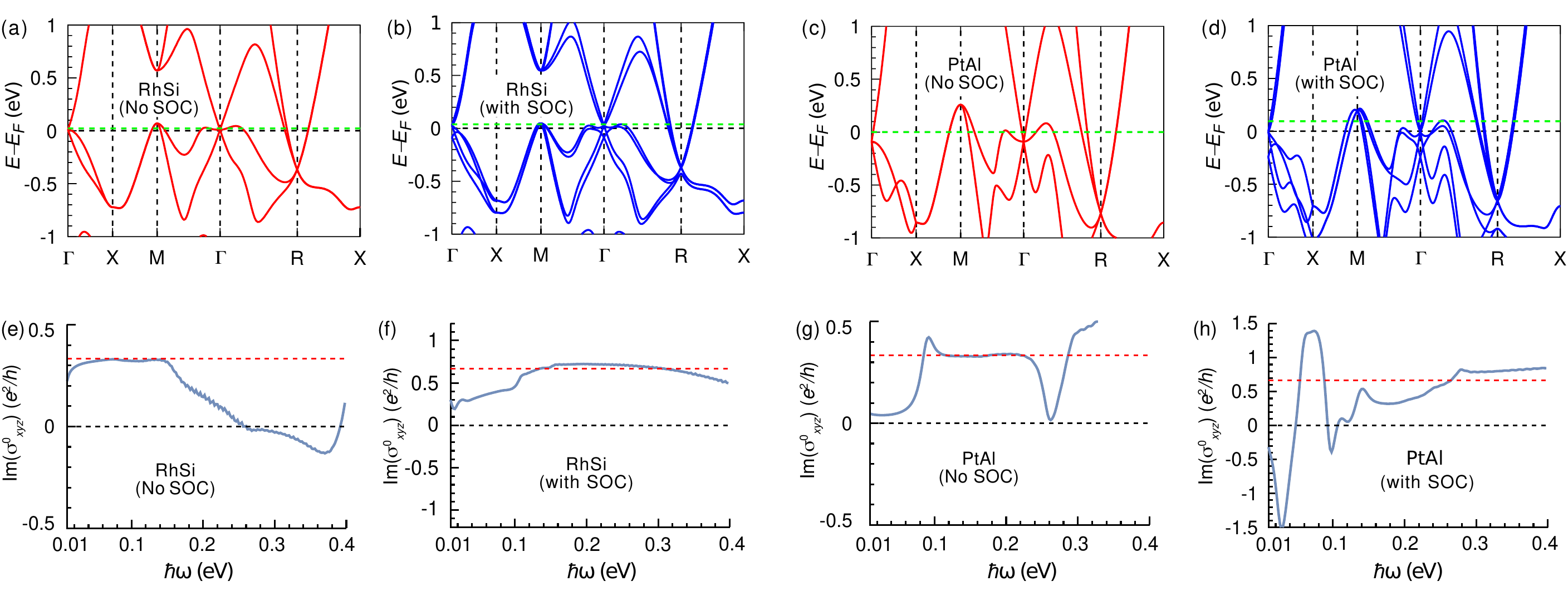}
\caption{
Interband circular dichroism of chiral semimetals RhSi and PtAl.
(a,b) Band structure of RhSi (a) without spin-orbit coupling (b) with spin-orbit coupling. (c,d) Band structure of PtAl (c) without spin-orbit coupling (d) with spin-orbit coupling. The horizontal dashed green line denote the position of the chemical potential used in the calculation. (e-h) The ${\rm Im}(\sigma^0_{xyz})$ for these four cases. Note that, in the absence of spin-orbit coupling, ${\rm Im}(\sigma^0_{xyz})$ per spin channel is quantized around $e^2/3h$ for both RhSi and PtAl. In the presence of spin-orbit coupling, the ${\rm Im}(\sigma^0_{xyz})$ deviates from the quantized value ($2e^2/3h$) due to large spin-orbit coupling. Interestingly, the deviation is still within $\sim10\%$ for RhSi, and it is within $\sim20\%$ for PtAl, despite significantly large spin-orbit coupling in these compounds in comparison to CoSi.}
\label{fig:PtAl}
\end{figure}

\begin{figure}[ht!]
\includegraphics[width=\textwidth]{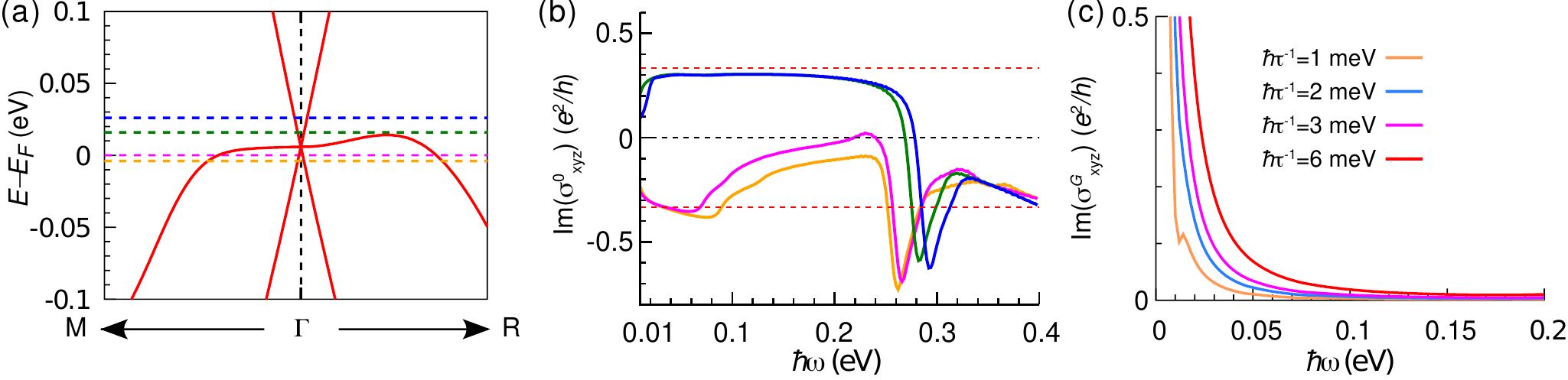}
\caption{
Dependence of circular dichroism on chemical potential and the broadening parameter in CoSi. (a) Band structure of CoSi without including the effect of spin-orbit coupling, zoomed in around the $\Gamma$ point. Horizontal dashed lines denote artificial shifts in the position of the chemical potential to simulate effect of doping. Such chemical doping effect has recently been experimentally achieved in RhSi~\cite{RhSi_doping}. (b) ${\rm Im}(\sigma^0_{xyz})$ for different chemical potentials  shown in (a). The color scheme is the same in panel (a) and (b). The sign of ${\rm Im}(\sigma^0_{xyz})$ is dependent on the energy position of the threefold degenerate point. (c) ${\rm Im}(\sigma^G_{xyz})$ for four different values of the broadening parameter, $\hbar\tau^{-1}$. Clearly, reducing $\hbar\tau^{-1}$ reduces the ${\rm Im}(\sigma^G_{xyz})$.}
\label{fig:mu}
\end{figure}

\end{widetext}

\end{document}